\def\presentation{
\voffset -.50in  \hoffset -.19in
\oddsidemargin 0in \evensidemargin 0in
\marginparwidth .75in \marginparsep 7pt \topmargin 0in
\headheight 12pt \headsep .25in
\footheight 18pt \footskip .35in
\textheight 9.5in \textwidth 6.5in
\columnsep 10pt \columnseprule 0pt }
\documentstyle{article}
\presentation
\begin{document}
%

%
\catcode`\@=11
\font\tenmsa=msam10 at 12truept
\font\sevenmsa=msam7
\font\fivemsa=msam5
\font\tenmsb=msbm10 at 12truept
\font\sevenmsb=msbm7 at 9truept
\font\fivemsb=msbm5 at 7truept
\newfam\msafam
\newfam\msbfam
\textfont\msafam=\tenmsa \scriptfont\msafam=\sevenmsa
 \scriptscriptfont\msafam=\fivemsa
\textfont\msbfam=\tenmsb \scriptfont\msbfam=\sevenmsb
 \scriptscriptfont\msbfam=\fivemsb

\def\hexnumber@#1{\ifcase#1 0\or1\or2\or3\or4\or5\or6\or7\or8\or9\or
	A\or B\or C\or D\or E\or F\fi }

\def\msb{\tenmsb\fam\msbfam}
\def\Bbb{\ifmmode\let\next\Bbb@\else
 \def\next{\errmessage{Use \string\Bbb\space only in math mode}}\fi\next}
\def\Bbb@#1{{\Bbb@@{#1}}}
\def\Bbb@@#1{\fam\msbfam#1}
\def\z{\Bbb Z}
\def\tilde{\widetilde}
\def\bar{\overline}
\def\hat{\widehat}
\def\*{\star}
\def\[{\left[}
\def\]{\right]}
\def\({\left(}
\def\){\right)}
\def\zb{{\bar{z} }}
\def\frac#1#2{{#1 \over #2}}
\def\inv#1{{1 \over #1}}
\def\half{{1 \over 2}}
\def\d{\partial}
\def\der#1{{\partial \over \partial #1}}
\def\dd#1#2{{\partial #1 \over \partial #2}}
\def\vev#1{\langle #1 \rangle}
\def\bra#1{{\langle #1 |  }}
\def\ket#1{ | #1 \rangle}
\def\rvac{\hbox{$\vert 0\rangle$}}
\def\lvac{\hbox{$\langle 0 \vert $}}
\def\2pi{\hbox{$2\pi i$}}
\def\e#1{{\rm e}^{^{\textstyle #1}}}
\def\grad#1{\,\nabla\!_{{#1}}\,}
\def\dsl{\raise.15ex\hbox{/}\kern-.57em\partial}
\def\Dsl{\,\raise.15ex\hbox{/}\mkern-.13.5mu D}
\def\comm#1#2{ \BBL\ #1\ ,\ #2 \BBR }
\def\x{\stackrel{\otimes}{,}}
\def\det{ {\rm det}}
\def\tr{{\rm tr}}
%
%
\def\th{\theta}		\def\Th{\Theta}
\def\ga{\gamma}		\def\Ga{\Gamma}
\def\be{\beta}
\def\al{\alpha}
\def\ep{\epsilon}
\def\la{\lambda}	\def\La{\Lambda}
\def\de{\delta}		\def\De{\Delta}
\def\om{\omega}		\def\Om{\Omega}
\def\sig{\sigma}	\def\Sig{\Sigma}
\def\vphi{\varphi}
%
%
\def\CA{{\cal A}}	\def\CB{{\cal B}}	\def\CC{{\cal C}}
\def\CD{{\cal D}}	\def\CE{{\cal E}}	\def\CF{{\cal F}}
\def\CG{{\cal G}}	\def\CH{{\cal H}}	\def\CI{{\cal J}}
\def\CJ{{\cal J}}	\def\CK{{\cal K}}	\def\CL{{\cal L}}
\def\CM{{\cal M}}	\def\CN{{\cal N}}	\def\CO{{\cal O}}
\def\CP{{\cal P}}	\def\CQ{{\cal Q}}	\def\CR{{\cal R}}
\def\CS{{\cal S}}	\def\CT{{\cal T}}	\def\CU{{\cal U}}
\def\CV{{\cal V}}	\def\CW{{\cal W}}	\def\CX{{\cal X}}
\def\CY{{\cal Y}}	\def\CZ{{\cal Z}}
%
%
\font\numbers=cmss12
\font\upright=cmu10 scaled\magstep1
\def\stroke{\vrule height8pt width0.4pt depth-0.1pt}
\def\topfleck{\vrule height8pt width0.5pt depth-5.9pt}
\def\botfleck{\vrule height2pt width0.5pt depth0.1pt}
\def\Zmath{\vcenter{\hbox{\numbers\rlap{\rlap{Z}\kern
		0.8pt\topfleck}\kern
		2.2pt \rlap Z\kern 6pt\botfleck\kern 1pt}}}
\def\Qmath{\vcenter{\hbox{\upright\rlap{\rlap{Q}\kern
                   3.8pt\stroke}\phantom{Q}}}}
\def\Nmath{\vcenter{\hbox{\upright\rlap{I}\kern 1.7pt N}}}
\def\Cmath{\vcenter{\hbox{\upright\rlap{\rlap{C}\kern
                   3.8pt\stroke}\phantom{C}}}}
\def\Rmath{\vcenter{\hbox{\upright\rlap{I}\kern 1.7pt R}}}
\def\Z{\ifmmode\Zmath\else$\Zmath$\fi}
\def\Q{\ifmmode\Qmath\else$\Qmath$\fi}
\def\N{\ifmmode\Nmath\else$\Nmath$\fi}
\def\R{\ifmmode\Rmath\else$\Rmath$\fi}
\def\cadremath#1{\vbox{\hrule\hbox{\vrule\kern8pt\vbox{\kern8pt
			\hbox{$\displaystyle #1$}\kern8pt} 
			\kern8pt\vrule}\hrule}}
\def\proof{\noindent Proof. \hfill \break}
\def\square{\hfill
\vrule height6pt width6pt depth1pt \\}
%
%
\def\debut{ \begin{eqnarray} }
\def\fin{ \end{eqnarray} }
\def\non{ \nonumber }
%

%
%
\rightline{SPhT-96-145}
\vskip 1cm
\centerline{\LARGE On Root Multiplicities of Some}
\bigskip
\centerline{\LARGE Hyperbolic Kac-Moody Algebras}
\vskip 1cm
\centerline{\large Michel Bauer and  Denis Bernard
\footnote[2]{Member of the CNRS} }
\centerline{Service de Physique Th\'eorique de Saclay
\footnote[3]{\it Laboratoire de la Direction des Sciences de la
Mati\`ere du Commisariat \`a l'Energie Atomique.}}
\centerline{F-91191, Gif-sur-Yvette, France.}
\vskip2cm
Abstract.\\
Using the coset construction, we compute the root multiplicities at
level three for some hyperbolic  
Kac-Moody algebras including the basic hyperbolic extension
of ${A_1^{(1)}}$ and $E_{10}$.
\vfill
\newpage
%
%
Hyperbolic or generalized Kac-Moody algebras \cite{kac} re-appear periodically
in string theory. There have been recent indications that they
should play a role in duality properties of supersymmetric
gauge theories \cite{harmoo} \cite{dvv}. It is indeed quite
striking that the modular group is often
a subgroup of the Weyl group of hyperbolic Kac-Moody algebras.
They also seem to have a role to play in two dimensional field
theory since some affine Toda field theories are known \cite{bl}
to be invariant under quantum affine algebras but with a central charge
acting as a topological charge. This suggests that hyperbolic algebras
could be related to the spectrum generating algebra of these models.

However little in known about hyperbolic Kac-Moody algebras,
(but more is known concerning generalized Kac-Moody algebras).
In particular the root multiplicities are known only for a small
subset of roots, the so-called level two roots \cite{FF83,kacwak}. The
aim of this 
short note is to shlightly extend this knowledge by determining
the root multiplicities for the level three roots.

\section{Hyperbolic extension of affine algebras.}

\subsection{Presentation.}
We are going to consider Kac-Moody algebras which are extensions
of affine algebras by a basic representation. We mainly  use notation from
the book \cite{kac}. We denote by $G_0$ the underlying affine algebra,
which we assume to be untwisted, and by $\hat G$ its extension.

Let $(a_{ij})$, $i,j=0,1,\cdots,r$, be the Cartan matrix of $G_0$,
and $e_i,f_i, h_i$ be the generators associated to the simple 
root $\al_i$. By convention the root $\al_0$ is the extended root
of the affine algebra $G_0$ and the remaining roots $\al_j$,
$j=1,\cdots,r$, are those of a finite simple Lie algebra, which 
we denote by $\CG$.  As usual, we extend $G_0$ by adding 
a derivation $d$  commuting with the Cartan 
generators $h_i$ and such that $[d,e_i]=-\de_{i,0}e_i$, and
$[d,f_i]=\de_{i,0}f_i$. In the following $G_0$ will refer to the
affine algebra extended by the derivation.
It will be useful to introduce a
basis $\{d;k;J^a_n, \; n\in \z, \; a=1,\cdots, dim~\CG\}$ of $G_0$ with
Lie brackets: 
\debut
[J^a_n,J^b_m] &=& f^{ab}_c J^c_{n+m} + nk q^{ab}\de_{n+m,0} \nonumber \\  
\[d,J^a_n\] &=& - nJ^a_n  \non
\fin
where $f^{ab}_c$ and $q^{ab}$ are the structure constants and
the Killing form of $\CG$, and $k$ is central. 
The invariant bilinear form of
$G_0$ is defined by $(J^a_n,J^b_m)=q^{ab}\de_{n+m,0}$ and
$(d,k)=-1$. As is well known, the derivation $d$ can be realized on
any integrable highest weight 
$G_0$-module as $d=L_0+ C^{st}$
where $L_0$ is the zero mode of the Sugawara Virasoro generators:
\debut
L_0 = \inv{2(k+h^*)}\sum_{a,b}\({ q_{ab}J^a_0J^b_0
+ 2\sum_{n>0} q_{ab} J^a_{-n}J^b_n }\) \non
\fin 
In the above formula, $h^*$ is the dual Coxeter number of $\CG$ and
$q_{ab}$ the 
inverse Killing form, $q^{ab}q_{bc} =\de^a_c$.

The Cartan matrix $\hat a$ of the extension $\hat G$ of $G_0$ is defined by~:
\debut
\hat a_{ij} &=& a_{ij},\quad for\quad i,j=0,1,\cdots,r \non\\
\hat a_{i,-1}&=& \hat a_{-1,i} = -\de_{i,0} \label{cartan}
\fin
The algebra $\hat G$ is thus of rank $(r+1)$. The root $\al_{-1}$
is usually referred as the overextended root.
Although this Cartan matrix always define a Kac-Moody
algebra, it does not always define a hyperbolic algebra. 
For instance it does not if the rank of $G_0$ is too big.
See ref.\cite{sac} for a complete classification of 
hyperbolic algebras.
The algebra $\hat G$ can be defined as the Lie algebra generated
by the elements $\{e_i,f_i,h_i, \; i=-1,0,1,\cdots,r\}$ with
relations:
\debut
[h_i,h_j]= 0  \quad &,& \quad \[e_i,f_j\] =\de_{ij} h_i \non\\
\[h_i,e_j\]= a_{ij} e_j \quad &,&\quad
[h_i,f_j]= -a_{ij} f_j \non\\
(ad e_i)^{1-a_{ij}} e_j & = &  (ad f_i)^{1-a_{ij}} f_j =0 \quad i \neq j \non
\fin
Notice that these relations in particular imply that we may identify
$d$ with $h_{-1}$ up to a multiple of $k$. Consistency of the
following constructions enforces the choice $d=h_{-1}+k$.

We will use an alternative presentation of $\hat G$ due to 
Feingold and Frenkel \cite{FF83}, see also \cite{kacwak} or \cite{FF93}.
It corresponds to
decompose $\hat G$ with respect to its affine subalgebra
$G_0$. This yields a graded decomposition of $\hat G$ as:
\debut
\hat G = \cdots + G_{-2}+G_{-1}+ G_0 +G_{1}+G_{2}+\cdots\non
\fin
where each subspaces $G_{n}$ are $G_0$-modules of level $n$.
So this decomposition is graded by $k$ the central element
of $G_0$.

By definition the modules $G_1$ (resp. $G_{-1}$) are the highest
(resp. lowest) $G_0$-modules generated by $f_{-1}$ (resp. $e_{-1}$).
They are isomorphic to the basic module $V(\La_0+\de)$ and
its dual $V^*(\La_0+\de)$:
\debut
G_1 \simeq V(\La_0+\de) \quad,\quad G_{-1}\simeq V^*(\La_0+\de) \non
\fin
The $d$-grading of the highest vector $\ket{\La_0+\de}$, 
which is identified with $f_{-1}$, is
fixed by the identification $d=h_{-1}+k$. It is
such that $d\ket{\La_0+\de}= - \ket{\La_0+\de}$.
We denote by $(v^*,v)$ for $v\in G_{1}$ and $v^*\in G_{-1}$
the pairing between $G_1$ and $G_{-1}$.

We first need to describe the Lie brackets between elements
in $G_0$ and $G_{\pm 1}$. Let $x\in G_0$, $v\in G_{1}$ and 
$v^*\in G_{-1}$, then by construction \cite{FF83}:
\begin{equation}
[x,v] = x\cdot v, \quad  [x,v^*] = x\cdot v^*, \quad \[v^*,v\] =
\sum_I~ (v^*,X^I\cdot v)~ X^I \label{commut} 
\end{equation}
where $x\cdot v$ and $x\cdot v^*$ refer to the action of
$x\in G_0$ on $v \in G_1$ or on $v^*\in G_{-1}$,
and where $(X^I)$ form an orthonormal basis of $G_0$.
It is easy to check that it satisfies the Jacobi identity:
\debut
[x,[v^*,v]]= [[x,v^*],v]+[v^*,[x,v]] \non
\fin

The Lie subalgebras $G_\pm= \sum_{n>0} G_{\pm n}$ are defined 
as certain quotients of the free Lie algebra $F_+=F(V)$ generated by $V$,
or $F_-=F(V^*)$ generated by $V^*$. A few basic facts
concerning free Lie algebras are recalled in the next subsection. 
Let us introduce $\hat F = F_- + G_0 + F_+$. As shown in \cite{FF83}
we can endow $\hat F$ with a Lie bracket which reduces to the
commutator (\ref{commut}) for the elements in $G_0,~G_{\pm 1}$.
Let now $J_\pm=\bigoplus_{n>0} J_{\pm n}$ be the ideals in $\hat F$
generated respectively by $J_{\pm 2}$ which are the subsets
of $F_{\pm 2}$ defined by:
\debut
J_{\pm 2} = \{ z \in F_{\pm 2} ; ~~[y,z] =0 ,
\forall\ y \in G_{\mp 1} \} \non
\fin
Let $J=J_+ + J_-$. It is an ideal in $\hat F$. 
As proved in \cite{FF83}, taking the quotient of $\hat F$
ensures that the Serre relations are fulfilled,
and the resulting algebra is isomorphic to
the hyperbolic extension $\hat G$:
\debut
\hat G \simeq \hat F/J \simeq F_-/J_- + G_0 + F_+/J_+ \label{defFF}
\fin
Ie. $G_\pm = F_\pm / J_\pm$.
This is the definition of $\hat G$ we will use to compute
the root multiplicities.

\subsection{Details on the free Lie algebra $F(V)$.}
We need a finer description of the free Lie algebra $F_+=F(V)$.
Let $(T(V), \otimes)$ be the tensor algebra of $V$. This is a graded
associative algebra, so we can endow it with a graded Lie algebra structure
by taking the usual commutator as Lie bracket. We identify $V$ with
the subspace of $T(V)$ consisting of elements of degree 1 and 
write $(T(V),[~,~])$ when we want to stress that we view $T(V)$ as a Lie
algebra. Let $F(V)$ be the smallest Lie subalgebra of $(T(V),[~,~])$
containing $V$. As $F(V)$ is generated by homogeneous elements of
degree 1, it is a graded Lie algebra, and we write $F(V)=
\bigoplus_{n=1}^{\infty} F_n$, with $F_1=V$. By definition, $V$ generates
$F(V)$. Let ${\cal U}(F(V))$ be the universal
envelopping algebra of $F(V)$. As usual we view $F(V)$ as a subspace
of ${\cal U}(F(V))$. Then it can be shown that \cite{Flie}
\proclaim Proposition.
i) The associative algebras $T(V)$ and ${\cal U}(F(V))$ are
canonically isomorphic.  \\
ii) Any linear map from $V$ to a Lie algebra $L$ extends uniquely to a Lie
algebra homomorphism from $F(V)$ to $L$. This  explains
why $F(V)$ is called a free Lie algebra. 
\par
\proof
The first part is a consequence of the universal properties of tensor
algebras and universal envelopping algebras. The identity map from $V
\subset T(V)$ to $V \subset {\cal 
U}(F(V))$ extends (in a unique way) to a homomorphism of associative
algebras. The identity map from $F(V) \subset {\cal U}(F(V))$ to $F(V)
\subset T(V)$, a Lie algebra homomorphism, extends (in a unique way)
to a homomorphism of associative 
algebras. The composition in any order of those two homomorphism is
the identity so they have to be isomorphisms. \\
The second part goes as follows : The linear map from $V$ to $L$
extends (in a unique way) to a homomorphism of associative algebras
from $T(V)$ to ${\cal U}(L)$. So by the first part there is a
homomorphims of associative algebras from ${\cal U}(F(V))$ to ${\cal
U}(L)$. This leads to a Lie algebra homomorphisms of the underlying
Lie algebras mapping $V$, hence $F(V)$ to $L$.
\square

Further information comes from the Poincar\'e-Birkhof-Witt (or PBW)
theorem which states that for a Lie algebra $L$, the graded algebra
associated to the filtered algebra ${\cal U}(L) =\cup_{n=0}^{\infty}
{\cal U}_n(L)$ is the symmetric algebra $S(L)=\bigoplus_{n=0}^{\infty}
S_n(L)$. In particular ${\cal U}_n(L)$ and $\bigoplus_{m=0}^{n}
S_m(L)$ carry isomorphic representations of $GL(L)$. For the free Lie
algebra $F(V)=\bigoplus_n F_n$ we see  
that $S(F(V)) \simeq \bigotimes_n S(F_n)$ and $T(V)$ are isomorphic as
graded representations of $GL(V)$. This isomorphism has several useful
manifestations, involving Poincar\'e series. 

We introduce a formal
variable $t$ to keep track of degrees. Then we can write
$T(V)=\frac{1}{1-tV}$ as a formal identity whose meaning is that the
expansion in $t$ on the right-hand side is:
\debut
T(V)=\frac{1}{1-tV}= 
1+tV+t^2V^{\otimes
2}+t^3V^{\otimes 3}+ \cdots \label{TV}
\fin
The corresponding formal expansion for $S(F_n)$ is 
$1+t^nF_n +t^{2n}S^2(F_n)+t^{3n}S^3(F_n)+\cdots$, which by a well-known
boson-fermion reciprocity can be rewritten as
\debut
S(F_n) =
\frac{1}{1-t^nF_n+t^{2n}\La^2(F_n)-t^{3n}\La^3(F_n)+\cdots} \label{SFn}
\fin
where $\La^p(F_n)$ denotes the wedge product of $p$ copies of $F_n$.
So we have 
\debut
\frac{1}{1 - t V} = \bigotimes_{n>0}\(\frac{1}{ 1 - t^n F_n + t^{2n} (\La^2F_n)
 - t^{3n} (\La^3 F_n) + \cdots }\) 
\fin
which we shall use by taking its inverse.

We also introduce a formal character. Let $\CA\in End(V)$ be
diagonalizable, with eigenvalues $a_\al$, treated as formal variables
of degree 1. It multiplicatively induces a degree preserving
endomorphism on $T(V)$, which we again denote by $\CA$. Let
$a_{\al_n}$ be the eigenvalues of the restriction of $\CA$ to $F_n$.
Then the character of $T(V)$ is $(1- t \sum_{\al}a_\al)^{-1}$ and
the character of $S(F_n)$ is $\prod_{\al_n}(1-t^n a_{\al_n})^{-1}$.
So we have
\debut
\(\inv{1- t \sum_{\al}a_\al } \)
=\prod_{n>0}\prod_{\al_n}\(\inv{1-t^n a_{\al_n}}\)
\fin
which we shall use by taking its logarithm. To summarize:

\proclaim Proposition.
Let $F(V)=\bigoplus_{n>0} F_n$ be the free Lie algebra generated by $V$,
then:\\
i) The spaces $F_n$ can be computed as subspaces of the tensor
algebra $T(V)$ using the following generating fonction:
\debut
1 - t V = \bigotimes_{n>0}\({ 1 - t^n F_n + t^{2n} (\La^2F_n)
 - t^{3n} (\La^3 F_n) + \cdots }\) \label{tauto}
\fin
ii) The characters are given by:
\debut
\log \(1- t \sum_{\al}a_\al \)= 
\sum_{n>0}\sum_{\al_n}\log \(1-t^n a_{\al_n}\)
\label{charac}
\fin
\par

The first formula (\ref{tauto}) gives for the first few spaces $F_n$:
\debut
F_2 &=& \La^2 V = V\wedge V \non\\
F_3 &=& V\otimes(V\wedge V) - \La^3 V \label{Fn}\\
F_4 &=& V\otimes F_3 - [ S^2(V\wedge V) - \Lambda^4 V ] \non
\fin
These formulas have a simple and nice interpretation:
all states in $F_n$ can be represented as nested commutators between
elements in $V$, and the relations (\ref{Fn})
describe the linear combinations between these states 
due to the Jacobi identity and the antisymmetry of the Lie bracket.
We can formulate this interpretation more precisely.
Consider the linear maps $I_{n+1}$ from $V\otimes F_n$ to $F_{n+1}$
defined by :
\debut
I_{n+1}(v\otimes f) = [ v, f] \label{defIn}
\fin
for any $v\in V$ and $f\in F_n$. By definition we have
the isomorphism:  
\debut
F_{n+1} \simeq (V\otimes F_n) / Ker I_{n+1} \label{Fcoset}
\fin
The relations (\ref{Fn}) tell us how to determine
the kernel of $I_{n+1}$ using the following sequences:
\def\longto{ \longrightarrow }
\debut
&&0 \longto \La^3 V \longto V\otimes(V\wedge V) 
{\buildrel  I_3  \over \longto} F_3 \longto 0 \non\\
&&0 \longto \Lambda^4 V \longto S^2(V\wedge V) 
{\buildrel id\otimes I_3 \over \longto}
V\otimes F_3 {\buildrel I_4 \over \longto} F_4 \longto 0 \non
\fin
It is easy to see that these sequences  define complexes.
The relations (\ref{Fn}) tell us that these complexes have
no cohomology, ie. the sequences are exact. Thus:
\debut
Ker I_3 &=& \La^3 V \label{kerI}\\
Ker I_4 &=& (id \otimes I_3) S^2(\La^2 V) \simeq S^2(\La^2V) - \La^4V\non
\fin 
This construction can clearly be generalized to any level.

The second formula (\ref{charac}) leads to
\debut
\(\sum_{\al}a_\al\)^n=\sum_{pq=n}p \(\sum_{\al_n}a_{\al_n}^q\)
\fin
For dimensions, this could be inverted using the
M\"obius function.
Anyway, the term $p=n$ involves the character $ChF_n=\sum_{\al_n} a_{\al_n}$
of the space $F_n$. For instance
\debut
ChF_2 &=& \half \({\sum_{\al} a_\al}\)^2 - \half 
\sum_{\al} a^2_\al \label{chf2}\\  
ChF_3 &=& \inv{3} \({\sum_{\al} a_\al}\)^3 - \inv{3}
\sum_{\al} a^3_\al \label{chf3}\\  
ChF_4 &=& \inv{4}\({\sum_{\al} a_\al}\)^4 - \inv{4}
\sum_{\al} a^4_\al -\inv{2}\sum_{\al_2} a^2_{\al_2}\label{chf4}
\fin

Finally let us point out that the algebra $\hat F = F_-+G_0+F_+$
is not a generalized Kac-Moody algebra \cite{borc}. In particular, taking the
quotient of $\hat F$ by the ideal generated by $J_{\pm 2}$ 
is required to have a contravariant form which is non
degenerate outside the Cartan subalgebra.

\subsection{Details on the ideal $J_2$}

To compute the roots multiplicities we need a finer identification
of $J_2$ and the ideal it generates. 
We describe how the Feingold-Frenkel presentation of the
hyperbolic algebras is related to the coset construction \cite{GKN}.
The fact that the coset construction should play a role
is clear from the formula (\ref{Fcoset}) which identifies
$F_{n+1}$ as a $G_0$-submodule of the tensor product of
$V$, a level one module, with $F_n$, a level $n$-module.

Let us introduce the tensor Casimir operator $\CC$ which is the
element of $G_0\otimes G_0$ defined by $\CC = \sum_I X^I\otimes X^I$, 
with $\{X^I\}$ an orthonormalized basis of $G_0$. In the basis 
$\{d;k;J^a_n, \; a=1,\cdots, dim\CG, \; n\in \z\}$ it reads:
\debut
\CC= -d\otimes k - k\otimes d + \sum_{n\in \z}
q_{ab} J^a_n \otimes J^b_{-n} 
\label{defCC}
\fin
Clearly, when acting on tensor product of two representations
$W_1$ and $W_2$ of $G_0$, the operator  $\CC$ commutes with 
the diagonal action of $G_0$. It is actually related to the coset 
Virasoro generators.

\proclaim Proposition.
Consider highest weight $G_0$-modules $W_1$ and $W_2$.
Suppose that on these modules we identify the derivation $d$
with $L_0 + \eta_{1,2}$ where $\eta_1$ and $\eta_2$ are constants. 
Then, on $W_1\otimes W_2$ we have~:
\debut
\CC= - k\otimes \eta_2- \eta_1 \otimes k - (\De(k) + h^*)L^{coset}_0
\label{Ccoset}
\fin
where $L^{coset}_0$ is the zero mode of the coset Virasoro generators~:
\debut
L^{coset}_0 = L_0\otimes 1 + 1\otimes L_0 
- \inv{2(\De(k) + h^* )} \sum_{a,b} \({ q_{ab}\De J^a_0\De J^b_0
+ 2\sum_{n>0} q_{ab}\De J^a_{-n}\De J^b_n }\) \label{l0coset}
\fin
where $\De(X)=X\otimes 1 + 1\otimes X$ for all $X\in G_0$.
\par

As a consequence, one gets the following description of $J_2$.
\debut
J_2 = Ker~\CC\big\vert_{V\wedge V} 
= \{ \ket{\om}\in V\wedge V ; \({L^{coset}_0\vert_{V\wedge V}}\)~ \ket{\om} 
= \frac{2}{h^*+2} \ket{\om} \}\label{Jdeux}
\fin
where $L^{coset}_0\vert_{V\wedge V}$ is the coset Virasoro
generator as defined in eq.(\ref{l0coset}) acting on $V\wedge V$.
The first equality, which follows from the definitions, was
proved in \cite{FF83}. The second equality follows from the
relation between $\CC$ and the coset Virasoro generators, from
the identification $d=L_0-1$ when acting on $V$, and from the
fact that $V\simeq V(\La_0+\de)$ is a level one module. 

Notice that the value $(\frac{2}{h^*+2})$ for $L^{coset}_0$ 
identifies the vectors of $J_2$ as highest weight vectors
of the coset Virasoro generators.

\subsection{ Roots multiplicities at level $2$ and $3$.}
In order to compute the roots multiplicities of $\hat G$
for the very first few levels, we extend the relation between $J_2$
and the coset construction to other components
$J_n$ of the ideal generated by $J_2$.

The definition of $\hat G =\bigoplus_{n\in \z} G_n$ 
as the quotient of $\hat F$ by the ideal 
$J=\bigoplus_{n\in \z}J_n$ generated by $J_{\pm2}$ gives us the
following description of $G_n$ for $n>0$ as~:
\debut
G_{n+1} \simeq F_{n+1}/ J_{n+1} \simeq 
V\otimes F_n - KerI_{n+1}- J_{n+1} \non
\fin
Using the formula (\ref{kerI}) for the kernel of the maps $I_{n+1}$,
we obtain~:

\proclaim Proposition.
\debut
G_2 &\simeq& (V\wedge V)- J_2 \non\\
G_3 &\simeq& V\otimes(V\wedge V) -\La^3 V - J_3 \quad with\quad
 J_3 \simeq V\otimes J_2- (V\otimes J_2)\cap \La^3V  \label{g3}\\
G_4 &\simeq& V\otimes F_3 - [S^2(V\wedge V)-\La^4V] - J_4 \quad with\quad
J_4 \simeq V\otimes J_3 - (V\otimes J_3)\cap Ker I_4 \non
\fin
\par

As we recalled above the $G_0$-module $J_2$ has been identified in \cite{FF83}.
Let us now identify the $G_0$-module $J_3$

\proclaim Proposition.
As subspaces of $V\otimes J_2$ we have the following inclusion,
\debut
(V\otimes J_2)\cap \La^3V \subset \{ \ket{\om} \in V\otimes J_2;~~
\({L_0^{coset}\vert_{V\otimes J_2}}\)~ \ket{\om} 
= \frac{4h^*+6}{(h^*+3)(h^*+2)} \ket{\om} \} \label{inclus}
\fin
where $L_0^{coset}\vert_{V\otimes J_2}$ is the coset Virasoro generator 
defined in eq.(\ref{l0coset}) acting on $V\otimes J_2$. 

\par
\proof
The space $V\otimes J_2$ is embeded into $V\otimes V\otimes V$.
Let $P_{ij}$ be the operator permuting the $i^{th}$ 
and $j^{th}$ copies of $V$ in $V^{\otimes^3}$.
Let us denote by $\CC_{ij}$ the tensor Casimir $\CC$ acting 
on the $i^{th}$ and $j^{th}$ copies of $V$ in $V^{\otimes^3}$.
If $\ket{\om}\in(V\otimes J_2)\cap \La^3V$ then $\CC_{23}\ket{\om}=0$,
by definition of $J_2$, and $P_{ij}\ket{\om}=-\ket{\om}$, by definition
of $\La^3V$. Therefore,
\debut
\ket{\om}\in(V\otimes J_2)\cap \La^3V \Rightarrow \CC_{ij}\ket{\om}=0
\quad \forall i,j=1,2~or~3 \non
\fin
since $P_{12}\CC_{23}= \CC_{13}P_{12}$, etc...
Now, taking into account the identification of $d=L_0-1$ in each
copy of $V$ in $V^{\otimes^3}$, we can express $\CC_{12}+\CC_{13}$
in terms of the coset Virasoro generators~:
\debut
\CC_{12}+\CC_{13}= -(h^*+3)L_0^{coset}\vert_{V\otimes J_2}
+(\frac{\CC_{23}}{h^*+2}) + (\frac{4h^*+6}{h^*+2}) \non
\fin
This proves the result .
\square
\par

\noindent Remark 1: this identifies a finite set of
possible $G_0$-modules, which are made of highest weight
vectors for $L^{coset}$ and which can be analysed case by case.

\noindent Remark 2: 
if $(V \otimes J_2)\cap \La^3V$ is the
trivial module $\{\ket{\om}=0\}$, then $J_3 \simeq V\otimes J_2 $,
and the subspace of level three $G_3$ is simply
\debut
G_3 \simeq V\otimes (V\wedge V) - \La^3V - (V \otimes J_2) \label{g3simple}
\fin
It is then a simple exercise in affine algebras to decompose this
space into irreducible $G_0$-modules, exercise that we will
describe in two cases in the next sections. This covers the case of
$\hat A_1^{(1)}$ 
but needs a slight generalization for $E_{10}$.

\noindent Remark 3: to compute the root multiplicities at level $4$
one would need to
compute the intersection $(V\otimes J_3)\cap Ker I_4$.
This is not so easy because the permutation group does
not act in a simple way on this space.

\section{Example $1$: $\hat{ A_1^{(1)} }$.}

In this section, we expand the formula (\ref{g3simple}) in the case of
$\hat{ A_1^{(1)} }$ which is the hyperbolic extension of the affine 
algebra $A_1^{(1)}$ by its basic representation.
Its Dynkin diagram is:

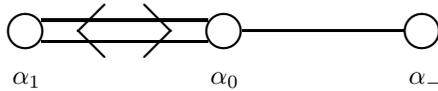
\centerline{\begin{picture}(200,25)(0,16)
\thicklines
\put(25,25){\circle{13}}
\put(100,25){\circle{13}}
\put(175,25){\circle{13}}
\put(31,29){\line(1,0){63}}
\put(31,21){\line(1,0){63}}
\put(107,25){\line(1,0){61}}
\put(45,25){\line(1,1){10}}
\put(45,25){\line(1,-1){10}}
\put(80,25){\line(-1,1){10}}
\put(80,25){\line(-1,-1){10}}
\put(20,5){$\alpha _1$}
\put(95,5){$\alpha _0$}
\put(170,5){$\alpha _{-1}$}
\end{picture}}
\par
\noindent It represents the Cartan matrix:
\debut
\hat a = \pmatrix{ 2& -2 & 0 \cr -2 & 2 & -1\cr 0 & -1& 2 \cr} \non
\fin
We label the roots by $\al_{1},\ \al_0$ and $\al_{-1}$ as indicated
on the Dynkin diagram. The dual Coxeter number is $h^*=2$.

The module $V$ is the basic level one
module $V(\La_0+\de)$. We recall that its character can be
expressed in terms the Theta and Dedekind functions~:
\debut
Ch~V \equiv Ch V(\La_0+\de)
= \frac{q^{-1}}{\eta(q)}~ \Theta_{\La_0} \label{charV}
\fin
Since $V$ has level one, all the coset constructions
associated to the identifications (\ref{Fcoset}) will
be related to the minimal representations of the Virasoro
algebra.

Consider first roots at level two. This space was already described
in \cite{FF83,kacwak} but we need to recall it for computing the
level three root multiplicities. The coset Virasoro associated
to $V\otimes V$ has central charge $c=\half$. Therefore
the decomposition of $V\otimes V$ with respect to the 
diagonal action of $A_1^{(1)}$ can be expressed in terms of
the $c=\half$ minimal Virasoro characters. As is well known one has~:
\debut
V\wedge V = V(\La_0+\de)\wedge V(\La_0+\de)
\simeq Vir_{c=1/2}^{h=1/2}(q)\otimes V(2\La_1 + 2\de) \non
\fin
where $Vir_{c}^{h}(q)$ denotes the Virasoro characters 
of the irreducible highest weight vector representation
of central charge $c$ and conformal weight $h$.
Since $h^*=2$, the ideal at level $2$ is 
\debut
J_2\simeq \ket{h=\half}\otimes V(2\La_1 + 2\de) \non
\fin
Therefore, we have \cite{FF83,kacwak}:
\debut
G_2 = \({ Vir_{c=1/2}^{h=1/2}(q) - q^{1/2} }\) \otimes V(2\La_1 + 2\de)
\label{g2su2}
\fin

Consider now the root multiplicites at level three. The Virasoro generators
representing the coset construction
associated to the isomorphism $F_3\simeq (V\otimes F_2)/Ker I_3 $
have central charge $c=7/10$. We need the decomposition of
the tensor product $V(\La_0+\de)\otimes V(2\La_1 + 2\de)$
with respect to the diagonal action of $A_1^{(1)}$. It is
given by~:
\debut
V(\La_0+\de)\otimes V(2\La_1 + 2\de) \simeq
Vir_{c=7/10}^{h=1/10}(q)\otimes V(\La_0+2\La_1 +3\de)
+ Vir_{c=7/10}^{h=3/2}(q)\otimes V(3\La_0+3\de) \label{cosetsu2}
\fin 
Comparing this formula and the inclusion (\ref{inclus}) 
we learn that $(V\otimes J_2)\cap \La^3V = {0}$ reduces to the
trivial module: $\frac{4h^*+6}{(h^*+3)(h^*+2)}=\frac{7}{10}$ and there is no
state $\ket{\om}\in (V\otimes J_2)$ 
such that $\({L_0^{coset}\vert_{V\otimes J_2}}\)~ \ket{\om}
= \frac{7}{10}\ket{\om}$.
Therefore 
\debut
J_3=V\otimes J_2 \quad for\quad \hat {A^{(1)}_1}\non
\fin
As a consequence we have~:
\debut
G_3&=& \[{Vir_{c=1/2}^{h=1/2}(q) - q^{1/2} }\]\otimes
 ~\[{Vir_{c=7/10}^{h=1/10}(q)\otimes V(\La_0+2\La_1 +3\de)
+ Vir_{c=7/10}^{h=3/2}(q)\otimes V(3\La_0+3\de) }\] \non\\
& & ~~~- \La^3V(\La_0+\de) \label{g3su2}
\fin
The last step consists in finding  an explicit 
expression for the character of $\La^3V(\La_0+\de)
\equiv \La^3V$.  This can be done using the formula (\ref{chf3}) 
and (\ref{charV}).  It gives~:
\debut
Ch(\La^3V) = \inv{3} \({Ch V }\)^3 
- \inv{3} \frac{q^{-3}}{\eta(q^3)} \Theta_{3\La_0} \label{chV3}
\fin
But $\({Ch V }\)^3$ is the character of $V^{\otimes^3}$ and
therefore can be expressed in terms of the Virasoro characters
as follows~:
\debut
\({ Ch V }\)^3 &=& \[{ Vir_{c=1/2}^{h=1/2}(q)Vir_{c=7/10}^{h=1/10}(q)
+ Vir_{c=1/2}^{h=0}(q)Vir_{c=7/10}^{h=3/5}(q) }\] ChV(\La_0+2\La_1+3\de) \non\\
&& ~~~+ \[{Vir_{c=1/2}^{h=1/2}(q)Vir_{c=7/10}^{h=3/2}(q) 
+ Vir_{c=1/2}^{h=0}(q)Vir_{c=7/10}^{h=0}(q) }\] ChV(3\La_0+3\de) \non 
\fin
The second term in (\ref{chV3}) is explicit but we can also 
reexpress it in terms of the $A^{(1)}_1$ characters of representations
of level $3$. To do it we use the fact that the Theta functions
of level three are linearly related to the characters at level three
with coefficients given by the so-called string functions:
\debut
Ch V(i\La_0+j\La_1) = \sum_{n,m} C^{nm}_{ij}(q) \Theta_{n\La_0+m\La_1} \non
\fin
At level three the string functions can again be expressed in
terms of the Virasoro characters but with central charge $c=4/5$.
Namely~:
\debut
C^{30}_{30}(q) &=& \inv{\eta(q)} 
\[{ Vir_{c=4/5}^{h=0}(q) + Vir_{c=4/5}^{h=3}(q) }\] \non\\
C^{12}_{30}(q) &=& \inv{\eta(q)}
\[{ Vir_{c=4/5}^{h=2/3}(q) }\] \non\\
C^{12}_{12}(q) &=& \inv{\eta(q)} 
\[{ Vir_{c=4/5}^{h=1/15}(q) }\] \non\\
C^{30}_{12}(q) &=& \inv{\eta(q)}
\[{ Vir_{c=4/5}^{h=2/5}(q) + Vir_{c=4/5}^{h=7/5}(q) }\] \non
\fin
Inverting the linear system relating the characters and the string
functions gives the expression of $\Theta_{3\La_0}$~:
\debut
\({C^{12}_{12}C^{30}_{30}-C^{30}_{12}C^{12}_{30}}\)(q)~\Theta_{3\La_0}
= C^{12}_{12}(q)~ Ch V(3\La_0) - C^{12}_{30}(q)~ Ch V(\La_0+2\La_1) \non
\fin
Gathering all the formula gives an explicit formula for the decomposition
of the level three space $G_3$ as a $G_0$-module. So it gives
an explicit formula for the root multiplicities at level three.
Explicit but ugly!

\section{Example $2$: $E_{10}$.}

In this section we present the root multiplicities at level
three for $E_{10}$. Once again it consists in making
formula (\ref{g3}) explicit. As we explained the 
$\hat {A_1^{(1)}}$ case in detail we will be more
sketchy for $E_{10}$.

The Dynkin diagram of $E_{10}$ is

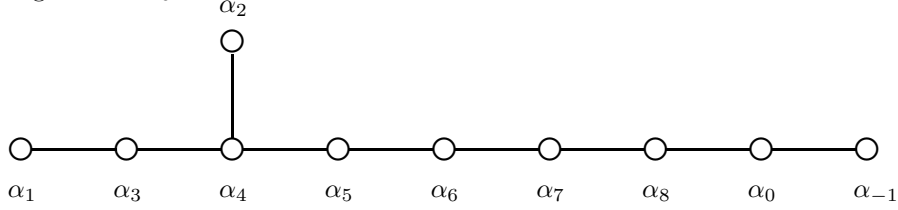
\centerline{\begin{picture}(360,80)
\thicklines
\multiput(25,25)(40,0){9}{\circle{8}}
\multiput(29,25)(40,0){8}{\line(1,0){32}}
\put(105,29){\line(0,1){32}}
\put(105,66){\circle{8}}
\put(20,7){$\alpha _1$}
\put(60,7){$\alpha _3$}
\put(100,7){$\alpha _4$}
\put(100,77){$\alpha _2$}
\put(140,7){$\alpha _5$}
\put(180,7){$\alpha _6$}
\put(220,7){$\alpha _7$}
\put(260,7){$\alpha _8$}
\put(300,7){$\alpha _0$}
\put(340,7){$\alpha _{-1}$}
\end{picture}}
\par
The indices refer to the root labelling. The dual Coxeter number is
$h^*=30$.
Let $\La_i$
be the weight dual to the affine root $\al_i$. The weight
$\La_0$ is the only integrable weight at level one for
the affine subalgebra $E_8^{(1)}$. The integrable weights 
at level two are $2\La_0$, $\La_8$ and $\La_1$.
Those integrable at level three are $3\La_0$, $\La_8+\La_0$,
$\La_1+\La_0$, $\La_2$ and $\La_7$.

The basic module which generates the hyperbolic extension of
$E^{(1)}_8$ is $V\simeq V(\La_0+\de)$.
As explained previously, the root multiplicities at level two
are related to the coset construction $(E^{(1)}_{8;k=1}
\otimes E^{(1)}_{8;k=1})/E^{(1)}_{8;k=2}$. This is
described by the $c=\half$ minimal Virasoro representation.
Explicitly,
\debut
V(\La_0+\de)\wedge V(\La_0+\de) \simeq Vir^{h=1/16}_{c=1/2}(q)
\otimes V(\La_8+2\de) \label{e8coset1}
\fin
Comparing with (\ref{Jdeux}) gives:
\debut
J_2 \simeq \ket{h=\inv{16}}\otimes V(\La_8+2\de) \label{e10j2}
\fin
Thus the root multiplicities at level two are \cite{FF83,kacwak}:
\debut
G_2 \simeq \[ Vir^{h=1/16}_{c=1/2}(q)-q^{1/16}\]\otimes V(\La_8+2\de)
\label{e10lvel2}
\fin
For future convenience we need the expression of the highest weight
vector $\ket{\La_8}$ in $J_2$~:
\debut
\ket{\La_8} = \({J_{-1}^{\th}\otimes 1 - 1\otimes J_{-1}^{\th}}\)
\ket{\La_0}\otimes\ket{\La_0} \non
\fin
where $\th$ is the highest root of $E_8$.

To compute the root multiplicities at level three we have
to evaluate $J_3$. The Virasoro generators
representing the coset construction associated to the isomorphism 
$F_3\simeq (V\otimes F_2)/Ker I_3 $
have central charge $c=1-\frac{6}{11.12}$, again a minimal model. One
computes that $\frac{4h^*+6}{(h^*+3)(h^*+2)}=\frac{21}{11.16}$. As
explained in the previous section, we first 
have to look for the state $\ket{\om}\in (V\otimes J_2)$ 
such that $ \({L_0^{coset}\vert_{V\otimes J_2}}\)~ \ket{\om}
= \frac{21}{11.16} \ket{\om}$. Contrary to the $\hat {A_1^{(1)}}$
case, for $E_{10}$ there is a candidate for such a state. It occurs in
the coset decomposition $V(\La_0)\otimes V(\La_8)/ V(\La_7)$.
More preciesly:

\proclaim Proposition.
For $E_{10}$ we have: 
\debut
(V\otimes J_2)\cap \La^3V \simeq \ket{h=\frac{21}{11.16}}
\otimes V(\La_7+3\de) \label{incluse10}
\fin
Here $\ket{h=\frac{21}{11.16}}$ denotes the highest weight
vector of the Virasoro coset algebra with conformal weight
$\frac{21}{11.16}$.
Thus, we have: $ J_3 \simeq (V\otimes J_2) - 
\ket{h=\frac{21}{11.16}}\otimes V(\La_7+3\de)$.
\par

\proof
The states $\ket{\om}$ which are highest weight vectors for the
affine algebra with weight $\La_7$ and for  the
Virasoro coset with conformal weight 
$\frac{4h^*+6}{(h^*+3)(h^*+2)}= \frac{21}{11.16}$,
occurs at level one in the tensor product $V(\La_0)\otimes V(\La_8)$.
Therefore, it has to be a linear combination of the four following
vectors:
\debut
\ket{\La_0}\otimes J_{-1}^\th J_0^{-\al_8} \ket{\La_8}
\quad &;& \quad 
J_{-1}^\th\ket{\La_0}\otimes J_0^{-\al_8} \ket{\La_8} \non\\
\ket{\La_0}\otimes J_{-1}^{\th-\al_8}\ket{\La_8}
\quad &;& \quad
J_{-1}^{\th-\al_8}\ket{\La_0}\otimes\ket{\La_8}\non
\fin
But recall that we have to view $\ket{\La_8}$ as the
element $\({J_{-1}^{\th}\otimes 1 - 1\otimes J_{-1}^{\th}}\)
\ket{\La_0}\otimes \ket{\La_0}$
in $V(\La_0)\otimes V(\La_0)$. Therefore, these four states can
written in terms of the action of $J_{-1}^\th$
and $J_{-1}^{\th-\al_8}$ on one of the three copies of $\ket{\La_0}$
in $\ket{\La_0}\otimes \ket{\La_0}\otimes \ket{\La_0}$. The only
state of this form which is in $\La^3 V$ is:
\debut
\ket{\om} &=& (J_{-1}^\th \otimes J_{-1}^{\th-\al_8} \otimes 1
-J_{-1}^{\th-\al_8} \otimes J_{-1}^\th \otimes 1
- J_{-1}^\th \otimes 1 \otimes J_{-1}^{\th-\al_8} \non\\
& & ~~~~~ + J_{-1}^{\th-\al_8} \otimes 1\otimes J_{-1}^\th 
+ 1 \otimes J_{-1}^\th \otimes J_{-1}^{\th-\al_8}
- 1 \otimes J_{-1}^{\th-\al_8} \otimes J_{-1}^\th )
\ket{\La_0}\otimes \ket{\La_0}\otimes \ket{\La_0} \non
\fin
It is easy to check that this is a highest weight vector
for the affine algebra. One may also verify that $\ket{\om} \in (J_2\otimes V)$,
since it can written alternatively as follows:
\debut
\ket{\om}= (\De J_0^{-\al_8}\ket{\La_8})\otimes (J_{-1}^\th \ket{\La_0})
+ (\De J_{-1}^{\th-\al_8}\ket{\La_8})\otimes \ket{\La_0}
-\ket{\La_8}\otimes(J_{-1}^{\th-\al_8}\ket{\La_0}) \non
\fin
The fact that the multiplicity
in eq.(\ref{incluse10}) is one follows from the fact that the
states $\ket{\om}$ defined above is unique.
This proves the result (\ref{incluse10}).
\square

As a corollary we obtain
the root multiplicities at level three for $E_{10}$:
\debut
G_3 \simeq V\otimes (V\wedge V) - \La^3V - (V\otimes J_2)
+\ket{h=\frac{21}{11.16}}\otimes V(\La_7+3\de)
\label{e10g3}
\fin
with $V=V(\La_0+\de)$ and $J_2\simeq \ket{h=\half}\otimes
V(\La_8+2\de)$.
It is a simple exercise, which we leave to the reader, to expand
this formula in terms of affine $E_8^{(1)}$ characters at level three
times branching functions. Since the coset construction $(E^{(1)}_{8;k=1}
\otimes E^{(1)}_{8;k=2})/E^{(1)}_{8;k=3}$ is a minimal model, the necessary
branching functions can be written in terms of characters of
$c=1-\frac{6}{11.12}$ Virasoro characters.

Clearly the same method can be applied to any extensions of an affine
algebra by one of its basic representations.

\end{document}